\documentclass[twocolumn,prd,aps,epsfig,floats,showpacs]{revtex4}
\def\DESepsf(#1 width #2){\epsfxsize=#2 \epsfbox{#1}}
\usepackage{epsfig}
\usepackage{graphicx}
\def\bmatrix{\left[\begin{array}}
\def\ematrix{\end{array}\right]}

\begin{document}

%

\let\a=\alpha      \let\b=\beta       \let\c=\chi        \let\d=\delta
\let\e=\varepsilon \let\f=\varphi     \let\g=\gamma      \let\h=\eta
\let\k=\kappa      \let\l=\lambda     \let\m=\mu
\let\o=\omega      \let\r=\varrho     \let\s=\sigma
\let\t=\tau        \let\th=\vartheta  \let\y=\upsilon    \let\x=\xi
\let\z=\zeta       \let\io=\iota      \let\vp=\varpi     \let\ro=\rho
\let\ph=\phi       \let\ep=\epsilon   \let\te=\theta
\let\n=\nu
\let\D=\Delta   \let\F=\Phi    \let\G=\Gamma  \let\L=\Lambda
\let\O=\Omega   \let\P=\Pi     \let\Ps=\Psi   \let\Si=\Sigma
\let\Th=\Theta  \let\X=\Xi     \let\Y=\Upsilon

%

%

\def\cA{{\cal A}}                \def\cB{{\cal B}}
\def\cC{{\cal C}}                \def\cD{{\cal D}}
\def\cE{{\cal E}}                \def\cF{{\cal F}}
\def\cG{{\cal G}}                \def\cH{{\cal H}}
\def\cI{{\cal I}}                \def\cJ{{\cal J}}
\def\cK{{\cal K}}                \def\cL{{\cal L}}
\def\cM{{\cal M}}                \def\cN{{\cal N}}
\def\cO{{\cal O}}                \def\cP{{\cal P}}
\def\cQ{{\cal Q}}                \def\cR{{\cal R}}
\def\cS{{\cal S}}                \def\cT{{\cal T}}
\def\cU{{\cal U}}                \def\cV{{\cal V}}
\def\cW{{\cal W}}                \def\cX{{\cal X}}
\def\cY{{\cal Y}}                \def\cZ{{\cal Z}}
%

\newcommand{\Ns}{N\hspace{-4.7mm}\not\hspace{2.7mm}}
\newcommand{\qs}{q\hspace{-3.7mm}\not\hspace{3.4mm}}
\newcommand{\ps}{p\hspace{-3.3mm}\not\hspace{1.2mm}}
\newcommand{\ks}{k\hspace{-3.3mm}\not\hspace{1.2mm}}
\newcommand{\des}{\partial\hspace{-4.mm}\not\hspace{2.5mm}}
\newcommand{\desco}{D\hspace{-4mm}\not\hspace{2mm}}


%
\title{\boldmath
Baryon Number Violation Involving Higher Generations }
\vfill
\author{Wei-Shu Hou$^{a,b}$}
\author{Makiko Nagashima$^a$}
\author{Andrea Soddu$^{a,c}$}
\affiliation{ $^a$Department of Physics, National Taiwan
 University, Taipei, Taiwan 106, R.O.C. \\
$^b$Stanford Linear Accelerator Center,
 Stanford, California 94309, U.S.A. \\
$^c$Department of Particle Physics, Weizmann Institute
 of Science, Rehovot 76100, Israel
}
%

%
%
\vfill
\begin{abstract}
Proton stability seems to constrain rather strongly any baryon
number violating process. We investigate the possibility of baryon
number violating processes involving right-handed dynamics or
higher generation quarks. Our results strongly suggest that there
will be no possibility to observe baryon number violation in
$\tau$ or higher generation quark decays, at any future machine.
\end{abstract}
%
\maketitle

\pagestyle{plain}

\section{Introduction}

Proton decay was suggested 30 years ago as a means to probe
physics at very high energy scale, such as~\cite{su5} Grand
Unified Theories (GUT). Baryon number violation is also a key
ingredient~\cite{Sakharov} to understand the baryon asymmetry of
our Universe. Experimental search for proton decay has
yielded~\cite{PDG} only limits so far, albeit very stringent ones,
and improvement requires huge underground facilities.

More recently, in an attempt to solve the hierarchy
problem~\cite{Arkani}, a new scenario has been proposed, which
assumes the existence of large compactified extra dimensions. It
has been observed that, having no test of gravity below the
millimeter scale, the radius of compactification can be much
larger than $M_{\rm Planck}^{-1}$, provided the extra dimensions
are accessible only to gravitational interactions. Standard Model
fields are instead localized on a three-dimensional wall whose
inverse thickness $L^{-1}$ can be of order TeV. Denoting by $V_n$
the volume of the extra space, with $n$ number of compact extra
dimensions, and by $M$ the fundamental scale of gravity, the
observed Planck mass is given by $M_{\rm Planck}^2=M^{n+2}V_n$. By
asking that $n \geq 2$, the fundamental scale $M$ could be close
to the electroweak scale.

In this new scenario, proton stability could be problematic. It is
not possible anymore to request that the gauge bosons which
mediate baryon number violating (BNV) processes to be super
massive~\cite{Perez}. A new mechanism has been suggested in
Ref.~\cite{Schmaltz} where leptons and quarks are localized at two
different positions in the extra space. The four-dimensional
effective coupling of the higher dimension operators responsible
for proton decay are then suppressed by the small overlap of the
quark and lepton wave-functions along the extra dimensions.

It has also been proposed~\cite{Schmaltz} that by localizing
different generations, and at the same time placing left and right
components at different positions along the extra space, one can
obtain in a natural way quark mass matrices which reproduce a
realistic mass spectrum and the quark mixing matrix. In
Ref.~\cite{Quiros} it has been shown that a split fermion
scenario, disfavored by constraints on flavor changing neutral
currents (FCNC) mediated by Kaluza-Klein gauge bosons, is still
possible if the three-dimensional wall has inverse thickness
$L^{-1} \geq 100$ TeV.

Having these scenarios in mind, it is reasonable to ask if BNV
processes involving higher generations, or right handed
components, could be faster than the processes involving the first
generation, which we know to be strongly suppressed. One can
imagine an extra space geography  where the second and third
generation quarks have big overlap with the leptons while the
first generation quarks are still sufficiently far away from the
lepton localization position.

It is therefore important to give an estimate of how fast BNV
processes involving higher generations could be, and how left or
right handed dynamics may affect them.
In Ref.~\cite{Marciano} it was suggested that one can estimate BNV
$\tau$ decay branching ratio by using proton decay constraints.
Applying the same approach, we estimate BNV processes involving
higher generations.

Let us write down all possible dimension six operators that
violate baryon number, but respect the symmetries of the Standard
Model (SM)~\cite{Weinberg,WZ}:
\begin{eqnarray}
O^{(1)}_{abcd} & = & (\bar{d}^c_{\alpha a R}u_{\beta b R})
(\bar{q}^c_{i \gamma c L}\ell_{j d L}) \, \epsilon_{\alpha \beta
\gamma}\epsilon_{ij}, \label{O1}\\* O^{(2)}_{abcd} & = &
(\bar{q}^c_{i\alpha a L}q_{j \beta b L}) (\bar{u}^c_{\gamma c
R}\ell_{d R}) \, \epsilon_{\alpha \beta \gamma}\epsilon_{ij},
\label{O2}\\* O^{(3)}_{abcd} & = & (\bar{q}^c_{i\alpha a L}q_{j
\beta b L}) (\bar{q}^c_{k \gamma c L}\ell_{l d L}) \,
\epsilon_{\alpha \beta \gamma}\epsilon_{ij}\epsilon_{kl},
\label{O3}\\* O^{(4)}_{abcd} & = & (\bar{q}^c_{i\alpha a L}q_{j
\beta b L}) (\bar{q}^c_{k \gamma c L}\ell_{l d L}) \,
\epsilon_{\alpha \beta \gamma} \nonumber \\* & & \hspace{3.cm}
\times(\vec{\tau}\epsilon)_{ij}(\vec{\tau}\epsilon)_{kl}\epsilon_{ik},
\label{O4}\\* O^{(5)}_{abcd} & = & (\bar{d}^c_{\alpha a R}u_{\beta
b R}) (\bar{u}^c_{\gamma c R}\ell_{d R}) \, \epsilon_{\alpha \beta
\gamma}, \label{O5}\\* O^{(6)}_{abcd} & = & (\bar{u}^c_{\alpha a
R}u_{\beta b R}) (\bar{d}^c_{\gamma c R}\ell_{d R}) \,
\epsilon_{\alpha \beta \gamma}, \label{O6}
\end{eqnarray}
where $\alpha, \beta, \gamma$ are color indices, $i,j$ are SU(2)
indices, $a,b,c,d$ are family indices, and $\epsilon_{\alpha \beta
\gamma}$, $\epsilon_{ij}$ are totally antisymmetric~\cite{foot}.
From Eqs. (\ref{O1})-(\ref{O6}) one can build the effective
interaction
\begin{equation}
\sum_{n=1}^6 C^{(n)}_{abcd}O^{(n)}_{abcd},
\end{equation}
by introducing  the coefficients $C^{(n)}_{abcd}$ of dimension
$d=-2$. The coefficients $C^{(n)}_{abcd}$ are model dependent
quantities which are sensitive to high energy scales. We shall
give a model independent estimate of the $C^{(n)}_{abcd}$
coefficients by using experimental limits on nucleon decays.

The structure of this paper is as follows. In section II we
estimate BNV $\tau$ decay branching ratios by using proton decay
constraints. In section III we apply the same approach to higher
generation quarks and consider BNV processes mediated by tree
diagrams. In section IV we show that loop diagrams give much
stronger constraints. In section V we estimate BNV $D$ and $B$
meson decays. Our conclusions follow.

\begin{figure}[tb]
\scalebox{0.4}{\includegraphics{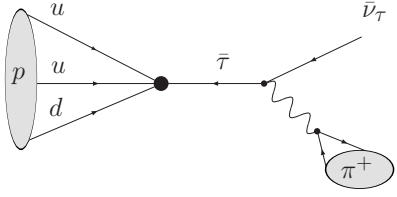}}
\caption{{\label{fig:ptonupiBNV}}$p\rightarrow \bar{\nu}_{\tau}
\pi^+$ generated by $C_{uud\t}^R$.}
\end{figure}

\section{BNV Operators and $\tau$ decays}

We start by analyzing BNV processes involving $\tau$, which are
intuitively very intriguing.
The fact that the proton cannot decay into $\tau$ but only into
$\nu_{\tau}$ implies that BNV decays involving left-handed $\tau$
are highly constrained, but for right-handed $\tau$, BNV processes
could be in principle faster. However, following
Marciano~\cite{Marciano}, by considering processes with virtual
$\tau$, as for example $p \rightarrow ``{\rm virtual}\, \tau"
\rightarrow \bar{\nu}_{\tau}\pi^+$ (Fig.~1), one can also strongly
constrain $\tau$ decays involving right-handed $\tau$.
So far the search for $\tau^-\to \bar{p}\pi^0$, $\bar{p}\eta$,
$\bar{p}\pi^0\pi^0$, $\bar{p}\pi^0\eta$,
$\bar{p}\gamma$~\cite{ARGUS,CLEO,BELLE} and
$\Lambda(\bar{\Lambda})\pi^-$~\cite{BELLE} decay modes has given
only upper limits.

\begin{figure}[tb]
\scalebox{0.4}{\includegraphics{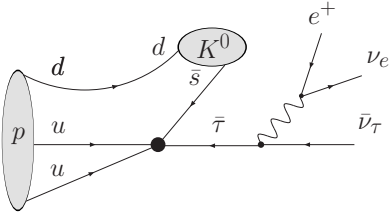}}
\caption{{\label{fig:tauKenunuBNV}}$p\rightarrow
K^0e^{+}\nu_{e}\bar{\nu}_{\tau} $ generated by $C_{uus\t}^R$.}
\end{figure}

Let us calculate explicitly the process of Fig.~1 when the BNV
vertex involves right-handed $\tau$. We use the chiral Lagrangian
obtained from the operators of Eqs.~(\ref{O2}), (\ref{O5}) and
(\ref{O6}). Each operator transforms under SU$_L(3) \times$
SU$_R(3)$ like $(\bar{3},3)$, hence the $\Delta B=1$ Lagrangian is
\begin{equation}
\alpha_p \left( C^R_{uud\tau}\, {\rm Tr}\, {\rm O} \xi^{\dag}
\bar{\rm\bf B}^c_R \xi^{\dag} + {C}^R_{uus\tau}\, {\rm Tr}\,
\tilde{\rm O} \xi^{\dag} \bar{\rm\bf B}^c_R \xi^{\dag}
\right)\tau_R, \label{Lchiral}
\end{equation}
where $\xi=\exp(i\,{\rm M}/f_{\pi})$ with $\rm M$ and ${\rm\bf B}$
the pseudo-Goldstone boson and octet baryon fields, respectively,
and $C^R_{uud\tau}$, ${C}^R_{uus\tau}$ are effective couplings for
operators in Eqs.~(\ref{O2}), (\ref{O5}) and (\ref{O6}). The
matrices $\rm O$ and $\tilde{\rm O}$ project out $\Delta S=0$ and
$1$ terms~\cite{CWH,Chada}.
The strong interaction parameter $\alpha_p$ is defined by the
relation $\langle 0 | \epsilon_{\alpha \beta \gamma} d_{\alpha R}
u_{\beta R} u_{\gamma L} |p \rangle \equiv \alpha_p\, u_{pL} $,
with $u_{p}$ the proton spinor. A recent lattice calculation
gives~\cite{Aoki} $\alpha_p =-0.015\ {\rm GeV}^3$, applicable for
proton decay. The BNV vertex in Fig.~1 corresponds to $\alpha_p
C_{uud\tau}^R \bar{\tau}^c_Rp_R$ from Eq.~(\ref{Lchiral}), with
$\alpha_p$ now at $\tau$ scale. A straightforward calculation for
the mode $p\rightarrow \bar{\nu}_{\tau} \pi^+$ shown in Fig.~1
gives the decay rate
\begin{equation}
\Gamma = \frac{\alpha_p^2 G_F^2 f_\pi^2 }{4
\pi}\frac{m_\tau^2(m_p^2-m_\pi^2)^2}
{m_p(m_\tau^2-m_p^2)^2}\left|C^R_{uud\tau}\right|^2,
\end{equation}
with $G_F$ the Fermi coupling constant. From the experimental
limit $\tau(p\rightarrow \bar{\nu} \pi^+) > 25 \times 10^{30}$
years, one can now constrain $C^R_{uud\tau}$ and obtain
\begin{equation}
C^R_{uud\tau}\lesssim 6.1 \times 10^{-18}\ {\rm TeV}^{-2},
\label{CRtau}
\end{equation}
where we have applied the same value for $\alpha_p$ as in proton
decay.

Using Eq.~(\ref{Lchiral}), the general SU$_L(3) \times$ SU$_R(3)$
invariant chiral Lagrangian involving the ${\rm M}$ and ${\rm\bf
B}$ fields~\cite{CWH}, and the $\tau$ width~\cite{PDG}, we obtain
the branching ratios,
\begin{eqnarray}
{\cal B}(\tau \to \overline p \pi^0) &\lesssim& 1.6\times10^{-4}\,
|C^R_{uud\tau}|^2, \label{tauppi} \\* {\cal B}(\tau \to \overline
p K^0) &\lesssim& 7.5\times10^{-5}\, | C^R_{uus\tau}|^2,
\label{taupK} \\* {\cal B}(\tau \to \overline\Lambda\pi^-)
&\lesssim& 1.3\times10^{-4}\, | C^R_{uus\tau}|^2, \label{tauLpi}
\end{eqnarray}
where $C_{uud\tau}^R$, $ C_{uus\tau}^R$ are in ${\rm TeV}^{-2}$
units.
Substituting Eq.~(\ref{CRtau}) in Eq.~(\ref{tauppi}) one obtains
\begin{equation}
{\cal B}(\tau \to \overline p \pi^0) \lesssim 5.9 \times 10^{-39}.
\end{equation}

The other two modes involving a strange quark depend on the
effective coupling ${C}_{uus\t}^R$, which corresponds to the
$-i\alpha_p/f_{\pi}( {C}_{uus\t}^R\bar{\tau}^c_Rp_R K^{0\dagger})$
term from Eq.~(\ref{Lchiral}), with $\a_p$ again at $\tau$ scale.
Considering the process $p \rightarrow K^0 e^+ \nu_{e}
\bar{\nu}_{\tau}$ of Fig.~2, and using $\tau
> 0.6 \cdot 10^{30}$ years for the semi-inclusive process
$p\rightarrow e^+ +$ anything, one obtains
\begin{equation}
{C}^R_{uus\tau}\lesssim 1.8 \times 10^{-13}\ {\rm TeV}^{-2},
\label{CRstau}
\end{equation}
where again we have applied the same value for $\alpha_p$ as in
proton decay. Substituting now Eq.~(\ref{CRstau}) in
Eq.~(\ref{taupK}) and Eq.~(\ref{tauLpi}) we also obtain
\begin{eqnarray}
{\cal B}(\tau \to \overline p K^0) &\lesssim& 1.4\times10^{-30} \,
, \label{taupKNUM} \\* {\cal B}(\tau \to \overline\Lambda\pi^-)
&\lesssim& 2.3\times10^{-30}\, . \label{tauLpiNUM}
\end{eqnarray}

One can generate $\tau \to \overline p \gamma$,
$\overline{\Sigma^+}\gamma$ decays from Eq.~(\ref{Lchiral}) by
radiating a photon off the external lines in the effective $\tau
\to \overline p$, $\overline{\Sigma^+}$ transitions, which are
analogous to $p\to \ell^+\gamma$. Direct radiation from the
transition vertex would involve higher dimension operators, hence
suppressed.
The radiative process should be down by a power of $\alpha_{\rm
em}/\pi$ compared to the modes we have just discussed. Indeed, we
find the branching fraction,
\begin{equation}
{\cal B}(\tau \to \overline p \gamma) \lesssim 2.8\times10^{-7}\,
|C_{uud\tau}^R|^2  \lesssim 1.0\times10^{-41}.
\end{equation}

We see that BNV $\t$ decays are extremely suppressed, and it is
unlikely that we will ever observe them at any future machine. We
now turn to BNV processes involving higher generation quarks.

\section{$C_{tt(c)b\ell}$ Coupling from $n\rightarrow \ell^+\pi^+\pi^-\pi^-$ }

\begin{figure}[tb]
\scalebox{0.4}{\includegraphics{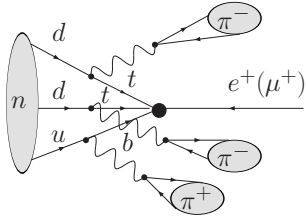}}
\caption{{\label{fig:nto3piep}}$n\rightarrow
\pi^+\pi^-\pi^-e^+(\mu^+)$ generated by $C_{ttb\ell}$.}
\end{figure}

In the previous section we showed how Marciano's suggestion of
considering virtual $\tau$ in proton decays can be used to
constrain $C_{uud\tau}^R$ and ${C}_{uus\tau}^R$. By analogy one
can then constrain any coupling $C_{abcd}$ through nucleon decays,
where now the family indices $a$, $b$, $c$ and $d$ can belong to
any generation.

As an example, we give in Fig.~3 the tree diagram for
$n\rightarrow \ell^+(p_1)\pi^+(p_2)\pi^-(p_3)\pi^-(p_4)$ induced
by a $C_{ttb\ell}$ coupling, where $\ell^+$ could be $e^+$ or
$\mu^+$. As internal states all possible combinations are allowed,
where a $b$ ($t$) quark propagator can be replaced by a $d$ or $s$
($u$ or $c$) quark. The amplitude will then be a combination of
terms each proportional to a different $C_{abcd}$.


As we are interested in giving only an order of magnitude
estimate, we can use $\tau(n\rightarrow e^+(\mu^+)+{\rm
anything})>0.6(12)\times 10^{30}\,{\rm years}$ to constrain the
$C_{abcd}$'s by assuming each time that only one single term
dominates. Taking for example the term proportional to
$C_{ttb\ell}$ as the dominant contribution and considering for
simplicity only the effect of $O^{(5)}_{ttb\ell}$, the amplitude
can be written as
\begin{eqnarray}
{\cal M}_{n\rightarrow \ell\pi\pi\pi}^{(5)} &=&
\frac{G_F^3f_{\pi}^3}{2\sqrt{2}} V_{ub}^* V_{td}^2 \;
C_{ttb\ell}^{(5)} \nonumber \\* & & \langle \ell^+ \pi^- \pi^+
\pi^-| \; [\bar{u}^c\ps_2(1+\g_5)b^c] \;
[\bar{b}^c_Rt_R\bar{t}^c_R \ell_R] \nonumber \\* & & \ \
[\bar{d}^c\ps_3(1+\g_5)t^c] \; [\bar{t}\ps_4(1-\g_5)d] \;
|n\rangle.
\end{eqnarray}
Using the matrix element $\langle 0 | \bar{u}^c\bar{d}^c d |n
\rangle \equiv \alpha_n\, \bar{v}_n $, we obtain
%
\begin{eqnarray}
i{\cal M}_{n\rightarrow \ell\pi\pi\pi}^{(5)}&=& 2\sqrt{2}\,G_F^3
f_{\pi}^3\, V_{ub}^*V_{td}^2 \, \a_n C_{ttb\ell}^{(5)}\, \nonumber
\\* & & \frac{1}{m_b m_t^2}\, {\rm tr}\left(\ps_2 \ps_4\right)\;
\bar{v}_n\ps_3(1+\g_5)v_\ell, \label{Mttbl}
\end{eqnarray}
where we have neglected the internal quark momenta $p_b$, $p_t$
and $p_t^{\prime}$ (all of $O(1\ \rm{GeV})$) with respect to
$m_b(1\ {\rm GeV})$ and $m_t(1\ {\rm GeV})$.

It is clear from Eq.~(\ref{Mttbl}) that the process $n\rightarrow
\ell^+\pi^+\pi^-\pi^-$ is highly suppressed, depending on the 6th
power of $G_F$ and the 18th power of the Cabibbo parameter $\l$
appearing in the Wolfenstein parametrization of the
Cabibbo-Kobayashi-Maskawa matrix (CKM). Further suppression comes
from four body phase space and the hadronic parameter $\a_n$.
Thus, the experimental constraint $\tau(n\rightarrow e^+(\mu^+) +
{\rm anything}) > 0.6(12)\times 10^{30}\,{\rm years}$ can be
satisfied without requiring the effective coupling
$C_{ttb\ell}^{(5)}$ to be very small. In fact $C_{ttb\ell}^{(5)}$
can be huge. We obtain for the $n\rightarrow
\ell^+\pi^+\pi^-\pi^-$ rate the following estimate
\begin{equation}
\G_{n\rightarrow \ell\pi\pi\pi}^{(5)}[{\rm TeV}] \sim 4\times
10^{-87}|C_{ttb\ell}^{(5)}|^2 \, ,
\end{equation}
which gives
\begin{equation}
C_{ttb\ell}^{(5)} \lesssim 10^{11}\, {\rm TeV}^{-2}.
\label{Cttbl5tree}
\end{equation}

Obviously, the upper limit in Eq.~(\ref{Cttbl5tree}) is much
higher than what could be realistic, but it would still seemingly
suggest that BNV processes like $t \rightarrow
\bar{b}\bar{b}W^-\ell^+$ could take place rather quickly. Assuming
now that $O_{tcb\ell}^{(i)}$ can give the dominant contribution,
from Eq.~(\ref{Mttbl}) it is clear that the constraint on
$C_{tcb\ell}$ is also very loose, and we obtain the rough estimate
\begin{equation}
C_{tcb\ell} \sim C_{ttb\ell} \frac{V_{td}}{V_{cd}}\frac{m_c}{m_t}
\lesssim 10^{7}\, {\rm TeV}^{-2}. \label{Ctcbl5tree}
\end{equation}
\noindent Decays like $t\rightarrow \bar{c}\bar{b}\ell^+ $ could
then easily compete with SM top decays making LHC a BNV machine.

Before getting carried away, we remind ourselves that the above
discussion of Fig.~3 is just an example. What we need to do is to
identify the fastest possible process generated by BNV involving
higher generation quarks. It is clear that we need to reduce the
powers of $G_F$ or $\lambda$. We now turn to what we feel is the
best process in the next section.

\begin{figure}[tb]
\scalebox{0.4}{\includegraphics{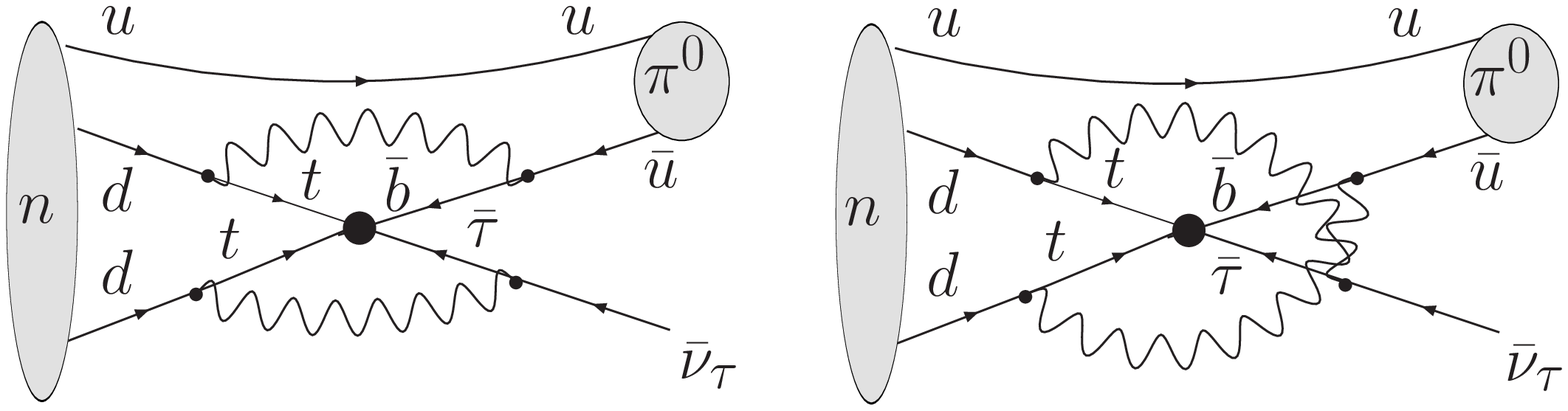}}
\caption{{\label{fig:ntonupiA}}$n\rightarrow \bar{\nu}_{\tau}
\pi^0$ generated by $C_{ttb\t}$.}
\end{figure}

\section{ $C_{tt(c)b\ell}$ Coupling from $n\rightarrow \bar{\nu}\pi^0$ }

The process $n\rightarrow \ell^+\pi^+\pi^-\pi^-$ in Fig.~3 is
highly suppressed, because for each tree level virtual $W$
emission, there is one power of $G_F$ suppression. It is clear
that a diagram involving $W$ emission at loop level will be less
suppressed. In Fig.~4 we give the two-loop radiative corrections
for the process $n\rightarrow \bar{\nu}_{\tau} \pi^0$ involving
$C_{ttb\tau}$; similar diagrams describe two-loop radiative
corrections for the process $n\rightarrow \bar{\nu}_{e,\mu} \pi^0$
with $\tau$ replaced by $e$ and $\mu$. The amplitude for the
process $n\rightarrow \bar{\nu}_{\tau} \pi^0$ at tree level is
proportional to $C_{udd\nu_{\tau}}$. When radiative corrections
are considered, $C_{udd\nu_{\tau}}$ mixes with all the other
couplings including $C_{ttb\tau}$. Knowing that
$C_{udd\nu_{\tau}}$ is already constrained from the tree level
calculation, we obtain a rough estimate of $C_{ttb\tau}$ by just
isolating the contribution to the amplitude given by the diagrams
of Fig.~4. We find in this way the most stringent constraint on
$C_{tt(c)b\ell}^{(i)}$. A very similar approach has been used in
Ref.~\cite{Wagner} to constrain the gauge couplings of an $SU(2)_1
\times SU(2)_2$ extension of the Standard Model. This theory can
in fact give rise to baryon and lepton number violating processes
via the instanton transition associated with the extended gauge
group.

Again we are interested only in order of magnitude results, and it
is therefore unnecessary to sum up all amplitudes coming from
different operators.
%
%
Let us assume for example that $C_{ttb\ell}^{(3)}$ gives the
dominant contribution
by considering only the effect of the operator $O^{(3)}_{ttb\ell}$
involving left-handed components. The amplitude for the first
diagram in Fig.~4 gives
\begin{eqnarray}
{\cal M}_{n\rightarrow \bar{\nu}\pi^0}^{(3)} &=& \frac{g^4}{64}
V_{ub}^* V_{td}^2 \; C_{ttb\ell}^{(3)} \; \int \frac{d^4 k_1 \;
d^4 k_2}{(2\pi)^4(2\pi)^4} \nonumber \\*
& & 
\frac{- g_{\m\n}g_{\a\b}}{(k_1^2-M_W^2)(k_2^2-M_W^2)} \; \langle
\bar{\n} \pi^0| \; [\bar{t}\g^\m(1-\g^5)d] \nonumber \\* & &
[\bar{d}^c \g^\a(1+\g^5)t^c] \; [\bar{u}^c \g^\b(1+\g^5)b^c]
\nonumber \\*
& & 
[\bar{e} \g^\n(1-\g^5)\n] \; [\bar{b}^c_L t_L \bar{t}^c_L \ell_L]
\; |n \rangle.
\end{eqnarray}
Using the matrix element $\langle \pi^0 | \bar{u}^c\bar{d}^c d |n
\rangle \equiv \alpha_{n\pi}\, \bar{v}_n $~\cite{fn1}, and after
some simplifications, we get
\begin{eqnarray}
{\cal M}_{n\rightarrow \bar{\nu}\pi^0} &=& -2\,G_F^2M_W^4\,
V_{ub}^*V_{td}^2\,\alpha_{n\pi} \,C^{(3)}_{ttb\ell} \nonumber \\*
& & \hspace{-14.mm}
 \int \frac{d^4 k_1 \; d^4 k_2}{(2\pi)^4(2\pi)^4} \;
  \frac{{\rm tr}\,\left[\g^\a \ps_b  \ps_t \g^\m
(1-\g^5)\right]}{(k_1^2-M_W^2)(k_2^2-M_W^2)} \nonumber \\* & &
\hspace{-14.mm} \frac{ \bar{v}_n \g_\a \ps_t^{\prime}\ps_l
\g_\m(1-\g_5)v_\n}{(p_b^2-m_b^2)(p_t^2-m_t^2)(p_t^{\prime
2}-m_t^2)(p_\ell^2-m_\ell^2)}, \label{IntO5}
\end{eqnarray}
where $k_1$ and $k_2$ are the momenta of the internal $W$ bosons,
and $p_b$, $p_t$, $p_t^{\prime}$ and $p_\ell$ are the momenta of
internal quarks and lepton.

Note that our BNV interactions are of four fermion interaction
form, hence not renormalizable. The two loops integral in
Eq.~(\ref{IntO5}) is ultraviolet divergent. To obtain a finite
result we can introduce a cut-off scale $\Lambda$ where we assume
our effective theory will break down. Setting all the external
momenta to zero we obtain
\begin{eqnarray}
{\cal M}_{n\rightarrow \bar{\nu}\pi^0} &=& -2\,G_F^2M_W^4\,
V_{ub}^*V_{td}^2\,\alpha_{n\pi} \,C^{(3)}_{ttb\ell} \nonumber \\*
& & \hspace{-14.mm} I^{(1)\r\s} I^{(2)\t\o}{\rm tr}\,\left[\g^\a
\g_\t \g_\r \g^\m (1-\g^5)\right] \nonumber \\* & &
\hspace{-14.mm} \bar{v}_n \g_\a \g_\o \g_\s \g_\m(1-\g_5)v_\n \,,
\label{IntO5fin}
\end{eqnarray}
where $I^{(i)\a\b}$ is defined as
\begin{equation}
I^{(i)\a\b} \equiv \int \frac{d^4k_i}{(2\pi)^4}\frac{k_i^\a
k_i^\b}{(k_i^2-m_a^2)(k_i^2-m_b^2)(k_i^2-M_W^2)},
\end{equation}
with $(m_a,m_b)$ respectively $(m_t,m_\ell)$ and $(m_b,m_t)$. The
integral $I^{(i)\a\b}$ has logarithmic dependence in the cut-off
scale $\Lambda$ plus corrections of order
$\Lambda^{-2}$~\cite{Wagner}. Setting the scale $\Lambda=10$ TeV,
and considering only the contribution to the amplitude coming from
$O^{(3)}_{ttb\ell}$, we obtain the following rough estimate for
the $n\rightarrow \bar{\nu}\pi^0$ rate
\begin{equation}
\G_{n\rightarrow \bar{\nu}\pi^0}^{(3)}[{\rm TeV}] \lesssim
10^{-38}|C_{ttb\ell}^{(3)}|^2 \,,
\end{equation}
where $C_{ttb\ell}^{(3)}$ is measured in ${\rm TeV}^{-2}$. Using
the experimental lower limit on the $n\rightarrow \bar{\nu}\pi^0$
partial lifetime, $\t_{n\rightarrow \bar{\nu}\pi^0}>112 \times
10^{30}$ years, we obtain
\begin{equation}
C_{ttb\ell}^{(3)} \lesssim 10^{-14}\ {\rm TeV}^{-2}.
\label{Cttbl3}
\end{equation}

One can follow the same approach for the operator
$O^{(5)}_{ttb\ell}$ which involves only right-handed
fields~\cite{fn2}, allowing for a weaker constraint on
$C_{ttb\ell}^{(5)}$
\begin{equation}
C_{ttb\ell}^{(5)} \lesssim 10^{-11}\ {\rm TeV}^{-2}.
\label{Cttbl5}
\end{equation}
Assuming now that $O_{tcb\ell}^{(i)}$ gives the dominant
contribution and using Eqs.~(\ref{Cttbl3}) and (\ref{Cttbl5}) we
obtain
\begin{equation}
C_{tcb\ell}^{(3)}\sim C_{ttb\ell}^{(3)}
\frac{V_{td}}{V_{cd}}\frac{I(m_b,m_t)}{I(m_b,m_c)} \lesssim
10^{-16}\ {\rm TeV}^{-2}, \ \
\end{equation}
with $I(m_a,m_b)$ defined by $I^{\a\b}(m_a,m_b) \equiv
g^{\a\b}I(m_a,m_b)$, and 
\begin{equation}
C_{tcb\ell}^{(5)}\sim C_{ttb\ell}^{(5)}
\frac{V_{td}}{V_{cd}}\frac{m_t}{m_c}\frac{I_S(m_b,m_t)}{I_S(m_b,m_c)}
\lesssim 10^{-11}\ {\rm TeV}^{-2}, \label{Ctcbl5}
\end{equation}
where
\begin{equation}
I_S \equiv \int
\frac{d^4k}{(2\pi)^4}\frac{1}{(k^2-m_a^2)(k^2-m_b^2)(k^2-M_W^2)}
\, .
\end{equation}
From Eq.~(\ref{Ctcbl5}) we expect $t\rightarrow
\bar{c}\bar{b}\ell^+$ can have at most a branching ratio
\begin{equation}
{\cal B}(t\rightarrow \bar{c}\bar{b}\ell^+) \sim
10^{-5}|C_{tcb\ell}^{(5)}|^2 \lesssim 10^{-27},
\end{equation}
where $C_{ttb\ell}^{(5)}$ is in ${\rm TeV}^{-2}$. This excludes
any possibility for discovering BNV top decays at future
colliders.

\section{$D$ and $B$ decays}

In this section we will give rough estimates for BNV $D$ and $B$
meson decay branching ratios. Following the arguments of the
previous sections for BNV $\tau$ and top quark decays we expect
BNV $D$ and $B$ processes to be also very suppressed. Nevertheless
we will still try to give some directions for a possible search.

We start by considering $D$ decays. We first note that, to allow
for a baryon in the final state, $\tau$ lepton is forbidden, hence
we consider only $c\to\ell^+\bar s\bar u$, $\ell^+\bar d\bar u$
processes with $\ell = e$, $\mu$. These lead to the decays
$D^+ \to \overline \Lambda\ell^+$ ($\overline\Delta^0\ell^+$ has
more background) and $D^0 \to \overline{\Sigma^+}\ell^+,\
\overline p\,\ell^+$. The $D^+$ mode has some advantage because it
has a longer lifetime and cleaner signature.
Since we are not concerned with hadronic effects, we emulate the
Lagrangian for $\tau$ decay, Eq.~(\ref{Lchiral}), which is itself
inspired by the chiral Lagrangian for proton decay. Assuming that
hadronic factors are of the same order of magnitude, we find the
Lagrangian for $D$ meson decays of interest,
\begin{eqnarray}
& & i\frac{\alpha_D}{f_D}  C_{ucd\ell}^RD^0 (\bar{p}^c_{R}
\ell_R)+ \nonumber \\*  && i\frac{\alpha_D}{f_D}  C_{ucs\ell}^R
\left( D^0 (\bar{\Sigma}^{+c}_{R} \ell_R) +
\sqrt{\frac{2}{3}}D^+(\bar{\Lambda}^{c}_{R} \ell_R) \right) .
\label{LchiralD}
\end{eqnarray}
From this we obtain the branching ratios,
\begin{eqnarray}
{\cal B}(D^+ \to \overline\Lambda \ell^+) &\lesssim&
3.9\times10^{-4}\, |C_{ucs\ell}^R|^2, \label{DLl} \\* {\cal B}(D^0
\to \overline{\Sigma^+} \ell^+) &\lesssim& 3.1\times10^{-4}\,
|C_{ucs\ell}^R|^2, \label{DSigl} \\* {\cal B}(D^0 \to \overline
p\ell^+) &\lesssim& 1.1\times10^{-4}\, |C_{ucd\ell}^R|^2,
\label{Dpl}
\end{eqnarray}
where the $C$'s are in units of TeV$^{-2}$, and we have set
$\alpha_D \lesssim \alpha_p$. These single baryon plus lepton
final states are rather distinct, but have so far never been
searched for. In fact, baryon number conservation instructs one to
consider baryon pair in final state, which is kinematically
forbidden. Thus, baryons in $D$ decay final states have commonly
been ignored.
$D$ decay to final states containing a baryon,
however, only probes baryon number violation in second and first
generations.

We do not consider $D_s$ decays, as it offers no special advantage
as a probe. We mention, however, the $\Lambda_c$ baryon, where the
process $\Lambda_c\to \tau^+K^0$ allows for $\tau$ in final state.
The process in principle probes $cus\to \tau^+$ interaction.
However, experimental study poses a problem. Not only would
$\Lambda_c$ be hadronically produced, having a $\tau$ in the final
state does not allow for full reconstruction. For this reason, we
shall drop $\tau$ from consideration even if it is kinematically
allowed in decay final state. For a rough estimate of the
branching ratios in Eqs.~(\ref{DLl}), (\ref{DSigl}) and
(\ref{Dpl}), we use for ${C}_{ucs\ell}^R$ and $C_{ucd\ell}^R$
respectively ${C}_{uus\tau}^R$ and $C_{uud\tau}^R$ from
Eqs.~(\ref{CRstau}) and (\ref{CRtau}) obtaining
\begin{eqnarray}
{\cal B}(D^+ \to \overline\Lambda \ell^+) &\lesssim&
1.3\times10^{-29}\, , \label{DLlestim} \\* {\cal B}(D^0 \to
\overline{\Sigma^+} \ell^+) &\lesssim& 1.0\times10^{-29}\, ,
\label{DSiglestim} \\* {\cal B}(D^0 \to \overline p\ell^+)
&\lesssim& 4.0\times10^{-39}. \label{Dplestim}
\end{eqnarray}


%



We now turn to the $B$ system. Having dropped $\tau$ from the
final state, the transitions of interest are $\bar b \to cc\ell$,
$cu\ell$, $uu\ell$, where $\ell = \mu$, $e$.
In a way, $\bar b\to \mu cc$ is analogous to $b\to \bar sss$, as
far as generation counting is concerned. This provides some extra
motivation, as CP violation in $B\to \phi K_S$ could be hinting at
presence of new physics. One can in principle have the double
charmed baryon $\Xi_{cc}$ in final state, i.e. $B^{0,+}\to
\Xi_{cc}^{+,++}\ell^-$.
The $\Xi_{cc}^+$ state is seen~\cite{Xicc} in $\Lambda_c^+
K^-\pi^+$ channel, although it still needs to be confirmed. By
analogy with $D^+\to K^-\pi^+\pi^+$ which has $\sim 9\%$ rate, we
expect ${\cal B}(\Xi_{cc}^+\to \Lambda_c^+ K^-\pi^+)$ could be of
order 20\%. Since the $cc$ system is generated from $\bar b \to
cc\ell$ transition with nonrelativistic motion, there should be
little suppression in the $B\to \Xi_{cc}$ form factor. Again, we
make analogy with $\tau\to \overline\Lambda\pi^-$, $D\to
\overline\Lambda\ell^+$ couplings of Eqs.~(\ref{Lchiral}) and
(\ref{LchiralD}), and estimate the branching ratio as,
\begin{equation}
{\cal B}(B^{0,+}\to\Xi_{cc}^{+,++}\ell^-) \lesssim 1.0 \times
10^{-4} \, |C^R_{ccb\ell}|^2.
 \label{BtoXiccl}
\end{equation}
The equivalent $\alpha_{B\to\Xi_{cc}}$ factor is set to $\alpha_p$
value. Using Eq.~(\ref{Ctcbl5}) we obtain
\begin{equation}
C_{ccb\mu}^{(5)} \sim  C_{tcb\tau}^{(5)}
\frac{V_{td}}{V_{cd}}\frac{m_t}{m_c}
\frac{I_S(m_t,m_\tau)}{I_S(m_c,m_\mu)} \lesssim 10^{-12}\ {\rm
TeV}^{-2}, \label{Cccbl5}
\end{equation}
and from Eqs.~(\ref{BtoXiccl}) and (\ref{Cccbl5}) we can estimate
${\cal B}(B^{0,+}\to\Xi_{cc}^{+,++}\ell^-) \lesssim 1.0 \times
10^{-28}$.

Although $\Xi_{cc}$ decay is likely dominated by only a few
channels, its reconstruction is still a problem. Typical
$\Lambda_c$ decay modes are a few percent, so one would be limited
by $\Lambda_c$ reconstruction efficiency. Perhaps one can also
consider double charm ($\Delta C = 2$) final states such as
$B^+\to D^{*+}D^0 p\ell^-$, which is quite close to threshold, but
would be stunning if seen: proton, $\ell^-$ and (slow) $\pi^+$
from $B$ vertex, while one has two $D$, rather than $D\overline
D$, tags. One could also consider the modes $\overline B\to
\Lambda_c^+D^{(*)0,+}\ell^-$. Again, it would be hard to set a
bound on the effective coupling, but can be searched for.

For $\bar b \to cu\ell$ generated processes such as $B^0\to
\Lambda_c^+\ell^-$, estimation becomes unreliable because of
relatively large energy release~\cite{HS}, and the $\Lambda_c$
would be in relativistic motion in the $B$ decay frame. But if we
continue with our approach, the branching ratio is,
\begin{equation}
{\cal B}(B^0\to \Lambda_c^+\ell^-) < 1.1 \times 10^{-4}\,
|C^R_{ucb\ell}|^2,
 \label{BtoLcl}
\end{equation}
where $\alpha_{B\to \Lambda_c} < \alpha_p$ is applied. The
suppression from $\alpha_{B\to \Lambda_c}^2/\alpha_p^2$ could be
as large as 100, if one compares the observed $\bar B^0 \to
\Lambda_c\bar p$ rate~\cite{Lcp} with that of $D^+\pi^-$. This
would give ${\cal B}(B^0\to \Lambda_c^+\ell^-) \lesssim 4 \times
10^{-30}$, if one applies the bound of Eq.~(\ref{CRstau}).

What may be more reasonable for the $B$ system, in view of the
large available energy, is to adopt an inclusive approach. Note
that $\bar b \to cu\ell^-$ bears some similarity with the usual
$\bar b\to \bar c\ell^+\nu$, except for the wrong charge
combination for $c\ell^-$ vs $\bar c\ell^+$ arising from $\bar b$
decay (plus replacing $\nu$ by $u$ quark). With $b$-tagging, one
searches for $B\to \Lambda_c^+\ell^-X$, with no missing energy. If
this can be properly achieved, it can be equated with the
inclusive $\bar b \to cu\ell^-$ branching ratio,
\begin{equation}
{\cal B}(\bar b\to cu\ell^-) = 4.0 \times 10^{-2}\,
|C^R_{ucb\ell}|^2,
 \label{btocul}
\end{equation}
and derive a more stringent bound. Applying the bound of
Eq.~(\ref{CRstau}), we find ${\cal B}(\bar b\to cu\ell^-) \lesssim
1.3 \times 10^{-27}$.
%
We remark that for $\bar b \to cu\ell^-$, because of unequal
flavor index, the $O^3$ operator of Eq.~(\ref{O3}) contributes.


For inclusive $\bar b\to uu\ell^-$ transitions, we get,
\begin{equation}
{\cal B}(\bar b\to uu\ell^-) = 7.4 \times 10^{-2}\,
|C^R_{uub\ell}|^2.
 \label{btouul}
\end{equation}
Exclusive search for $B\to p\ell^-$ can be done, but would not be
profitable as this mode would be highly form factor suppressed.
This is because of the large energy release~\cite{HS} and highly
relativistic proton motion. However, what one should search for is
$B\to p\ell^-X$ with $b$-tagging, where $p\ell^-$ has the wrong
charge combination and come from the $B$ decay vertex. Charm veto
for semileptonic decays should be applied. The study would
probably be background limited due to mistag and vertex
resolutions, but certainly should be pursued. The limit from
Eq.~(\ref{btouul}), after applying the bound of
Eq.~(\ref{CRstau}), is $2.4 \times 10^{-27}$.

\section{Conclusion}

We have used experimental data on nucleon decays to give an
estimate of BNV couplings involving higher generation leptons and
quarks. Using an approach very similar to the one proposed by
Marciano to infer very small branching ratio for BNV $\t$ decays,
we find that BNV processes involving higher generation quarks are
also extremely suppressed.

Despite our findings, we believe it is still worthy to look for
BNV processes in $\t$, charm, $B$, and maybe in the future in top
decays. Although we find that all BNV processes involving $\t$ and
higher generation quarks are too strongly suppressed by proton
stability, such that there seems no hope for observation at any
future machine, redundancy is very important. In this vein, we
note that the CLEO Collaboration performed $\tau^-\to
\bar{p}\pi^0$, $\bar{p}\eta$, $\bar{p}\pi^0\pi^0$,
$\bar{p}\pi^0\eta$, $\bar{p}\gamma$~\cite{CLEO} search {\it after}
the remark made by Marciano. The search has been extended recently
to $\tau^-\to\Lambda(\bar{\Lambda})\pi^-$~\cite{BELLE} by the
Belle Collaboration.

\vskip 0.3cm \noindent{\bf Acknowledgement}.\ \ This work is
supported in part by NSC-94-2112-M-002-035, NSC94-2811-M-002-053
and HPRN-CT-2002-00292. WSH thanks SLAC Theory Group for
hospitality. AS would like to thank Hsiang-nan Li and Diego
Guadagnoli for very useful discussions.


\begin{thebibliography}{99}
\bibitem{foot} All interactions in Eqs. (\ref{O1})-(\ref{O6}) conserve the
difference of baryon and lepton number $B-L$, which implies that
nucleons can decay only to antileptons but not leptons.
\bibitem{fn2}
Because of the different chiral structure of
the BNV operator $O^{(5)}$ with respect to $O^{(7)}$, the
amplitude is finite and no cut-off is required.
\end{thebibliography}
\end{document}